\documentclass[preprint]{aastex}
\usepackage{longtable}
\usepackage{color}
\usepackage{natbib}
\everymath{\rm}
\everydisplay{\sf}
\begin{document}

\title{Metal-Rich PNe in the Outer Reaches of M31}
\author{B. Balick}
\affil{Department of Astronomy, University of Washington, Seattle, WA 98195-1580, USA}
\email{balick@uw.edu}

\author{K.B. Kwitter}
\affil{Department of Astronomy, Williams College, Williamstown, MA 01267}
\email{kkwitter@williams.edu}

\author{R.L.M. Corradi\altaffilmark{a}}
\affil{ Instituto de Astrof{\'i}sica de Canarias, E-38200 La Laguna, Tenerife, Spain}
\email{rcorradi@iac.es}

\and
\author{R.B.C. Henry}
\affil{H.L. Dodge Department of Physics \& Astronomy, University of Oklahoma, Norman, OK 73019}
\email{henry@ou.edu}

\altaffiltext{a}{Departamento de Astrof{\'i}sica, Universidad de La Laguna, E-38206 La Laguna, Tenerife, Spain}

\begin{abstract}
Spectroscopic data of two relatively [O~III]-luminous -PNe have been obtained with the 10.4-m Gran Telescopio Canarias. M174 and M2496 are each $\sim$1 degree from the center of M31 along opposite sides of its minor axis.  The ensemble of these two distant PNe plus 16 similarly luminous outer-disk PNe published earlier (Kwitter et al. 2012 \& 2013) forms a homogeneous group in  luminosity, metal content, progenitor mass, age, and kinematics.  The main factual findings of our work are: (1) O/H (and other low-mass alpha elements and their ratios to O) is uniformly solar-like in all 18 PNe $<$12+log(O/H)$> = 8.62\pm0.14$); (2) the general sky distribution and kinematics of the ensemble much more closely resemble the rotation pattern of the classical disk of M31 than its halo or bulge; (3) the O/H gradient is surprisingly flat beyond R$_g$ $\sim$30 kpc, and may be flat throughout the entire range of R$_g$ covered in the full study.  The PNe are too metal-rich to be {\it bona fide} members of M31's disk or halo, and (4) the abundance patterns of the sample are distinct from those in the spiral galaxies M33, M81, and NGC~300.  Using standard PN age diagnostic methods (which are readily challengeable) we suggest that all of the PNe formed $\sim2$ GY ago in a starburst of metal-rich ISM that followed an M31--M33 encounter about 3 GY ago.  We review supporting evidence from stellar studies.  Other more prosaic explanations, such as dwarf galaxy assimilation, are unlikely.
\end{abstract}


\section{Introduction}

The cumulative history of metal enrichment in disk-dominated galaxies can be gleaned from the abundances of $\alpha$ elements of their ionized nebulae and young stars.  Tactically, ionized nebulae are particularly advantageous since their lines of the lighter $\alpha$ elements (notably oxygen, neon, argon, and sulfur) elements are luminous over a wide range of $\alpha$-element abundances (up to 10\% of the central star's total luminosity) and their emission-line fluxes are concentrated in narrow lines.  Absorption lines of B-type stars and Cepheids can also be used to estimate $\alpha$-element abundances and their radial gradients in nearby spirals, though the observations are more challenging, especially when the metal abundances are small.  

Until recently most studies of the  $\alpha$ elements and their spatial distributions have been confined to the Milky Way.  See \citet{H10} and Kwitter \& Henry (2012, ``KH12'') for a detailed review of the elemental abundances and their gradients measured in the Milky Way and its companion galaxies. However, the Milky Way presents several practical challenges.  Interstellar absorption and extinction limit the spatial range and spectral coverage of such studies to a few kpc near the Sun (except towards the Galactic anticenter).  Moreover, distances -- and hence the Galactic locations -- of many nebulae are poorly determined.  

For these reasons most recent studies of the distributions of the $\alpha$-element abundances have focused on nearby galaxies, with emphasis on those in the Local Group.  Although the emission-line luminosities of H~II regions are much larger then those in planetary nebulae (``PNe''), compact PNe offer several important strategic and tactical advantages for studies of metal enrichment in galaxies out to distances of $\sim 5$ Mpc: (1) unlike H~II regions and early-type stars PNe can be found in all types of galaxies and stellar environments; (2) PNe can be traced to much larger radii than H~II regions in disk-dominated galaxies ; (3) PNe are not constrained to spiral arms where local and foreground extinction hinder observations of important ultraviolet lines; and (4) the surface brightnesses of the most luminous PNe are far higher than those of H~II regions, which enables flux-limited observations of faint emission lines using the small entrance apertures generally needed for spectroscopy. 

Most nebular emission-line studies use O/H as a proxy for the $\alpha$ elements in the ISM since like other light $\alpha$ elements, oxygen is not altered by the central stars of H~II regions and PNe.  Thus the observed $\alpha$-element abundances characterize the ISM from which the stars formed.  In practice, the O/H ratio is the best determined of the light $\alpha$ elements owing to its prominent lines of O$^{+}$ and O$^{++}$ as well as to the reliability of measuring the  nebular temperature and correcting for unseen higher ionization states.   

Observations of PNe allow the radial distribution of O/H to be traced throughout large galaxies, including their innermost and outermost regions (bulges and halos) where star formation has long ceased.  \citet{JC99} used PNe to probe the abundances of M31 PNe, most of them in the bulge.  Magrini, Stanghellini, \& Villaver (2009, ``Mag09''), Stanghellini et al. (2010, ``Stang10''), and Stasi{\'n}ska et al. (2013, ``Stas13'') studied the metal gradients of PNe in M33, M81, and NGC~300, respectively.  We shall discuss their results later.  All of these galaxies show a systematically negative slope of the radial O/H gradient.  This is expected since oxygen enrichment is much stronger in the inner spiral-arm-dominated regions of disks where massive stars have been forming and expelling light $\alpha$ elements at the ends of their brief lives.  
 
Until recently systematic studies of the O/H distribution of M31 were based on spectroscopy of its prominent H~II regions.  Kwitter et al. (2012, ``Paper~I'') concentrated on the O/H abundances of PNe beyond the spiral arms.  They found that unlike the other nearby spiral galaxies, the PNe of M31 show a very shallow radial O/H gradient over $18 \la $ R$_g \la 45$ kpc; that is, to distances well beyond $R_{25}$\footnote{We adopt R$_{25}$=30.1 kpc, estimated from Fig.~1 of Courteau et al. (2011, ``Cour11''); see also Chapman et al. (2006,``Ch06''.} and the well-known warp in H~I (\citealt{B09,Chem09,Corb10}).  Their results show that O/H is within about 0.1 dex of its solar value at the outermost radius---surprisingly high since the underlying stellar population is generally characterized by [Fe/H] $\la-0.7$, but with some prominent exceptions discussed below. 

Here we report the results of abundance studies of two additional bright PNe, M174 and M2496, located about a degree from the nucleus along opposite sides of M31's minor axis.  These PNe, both found in the  extensive PN discovery survey of Merrett et al. (2006, ``M06''), lie along the edge of what appears to be a faint extended disk that shares the orientation and general rotation pattern of M31's inner disk (Ibata et al. 2005; see esp. their Fig.~1), but not in any of M31's known stellar streams  (Lewis et al. 2013, ``LB13'').  If the extended disk has the same inclination as the inner one then M174 and M2496 are at deprojected distances R$_g$ greater than 55 kpc.    

Our primary goals are (1) to measure the O/H values of M174 and M2496, (2) to see whether the O/H trends found in Paper~I continue to larger distances or whether PNe far beyond R$_{25}$ have the metal abundances of the halo as one might normally expect  (Cour11), and (3) to use these results to illuminate the evolution of M31's outermost structure and chemistry. The new observations are described in section 2, the observational results are presented in section 3, and we discuss these results within the context of complementary published results  for the same general regions of M31 and other spirals in section 4.  In section 5 we propose and evaluate a formation scenario for the ensemble of sixteen PNe observed in Paper~I and the two PNe reported here.

\section{Observations}
We present new spectrophotometry of two outer PNe in M31 from the catalogue of Merrett et al. (2006, ``M06''),  M174 and M2496 (Fig.~1a).  Their fundamental properties are presented in Table~1.  The values of R$_g$ were estimated using the method described in Paper~I and its erratum, (Kwitter et al. 2013). We also show the apparent magnitude in [O~III], m$_{5007}$, listed by M06.  M2496 is one of the brightest PNe in the M06 survey, and indeed, all 18 of our objects are within 0.7 magnitudes of the bright-end cutoff of the Planetary Nebula Luminosity Function (``PNLF''; see, e.g., \citealt{C10}).

The spectra were obtained on different nights in 2012 August and September at the 10.4-m Gran Telescopio Canarias (GTC) at the Roque de los Muchachos Observatory on the island of La Palma, Spain.  Finding charts were generated from SDSS images (DR8).  The GTC calibration plan provides, in each observing night, at least one spectrophotometric standard to be used for flux calibration.  The OSIRIS instrument was used in its longslit mode. The combination of grism R1000B and a slit width of 0.8\arcsec~provides a spectral dispersion of 2.1 \AA~per (binned 1x2) pixel, a resolution of 6.3 \AA, and a spectral coverage from 3700 \AA~ to 7850 \AA.  Seeing was 0.8\arcsec, and the spatial scale along the slit is 0.254\arcsec~per binned pixel.  Neither of the targets was spatially resolved.  Total exposure was 117 min for PN174 and 106 min for PN2496, each split into 4 sub-exposures. These data were obtained under photometric weather conditions and the slit was placed along the parallactic angle in order to minimize the effects of atmospheric dispersion.  The nebular data and results are given in Appendix~A.

An additional 145 min of exposure of PN174 were taken during a night with strong dust pollution from the Sahara desert. This, together with the observations of a poorly sampled spectrophotometric standard, prevented a precise flux calibration and therefore these data were not used for the nebular analysis.  However, they were added to the best-quality data for PN174 to improve the S/N of the faint stellar features discussed in Appendix~B.

\section{Results}
  
The observations and results for M174 and M2496 are compiled in Appendix A. The data analysis methodology is the same as in Paper I, so the results are consistent with the other PNe labeled in Fig.~1a. Line fluxes (Table~A1) are shown relative to H$\beta$=100 both before and after reddening corrections are applied. Table~A2 contains the ionic abundances for each of the PNe, listing the temperature (from Table~A3) used to calculate them. Also included for each element is the derived ionization correction factor ({\it icf}), calculated as described in \citet{KH01}. Table~A3 lists all the available diagnostics and total derived abundances for each PN. Estimated uncertainties are provided with the results. Table~A3 also lists corresponding solar abundances for comparison.

We used {\it cloudy} version c08.00 \citep{F98} with the measured nebular abundances to fit the observables -- the [O~III] luminosity found by M06, along with important emission line strengths and ratios -- by appropriate choices of initial conditions including  stellar surface gravity, luminosity, and temperature, and the nebular density structure.  On an H--R diagram the derived properties of the central stars place them near the evolutionary tracks for central stars of mass 0.6 M$_{\odot}$ and independently confirm the initial masses of the central stars near 2 M$_{\odot}$, as expected\footnote{We have accounted for extinction/reddening in terms of the observed vs. expected Balmer decrement in the nebular gas; we cannot measure and have ignored circumstellar extinction/reddening in this analysis.  Corrections for circumstellar dust would increase estimates of stellar temperature and luminosity, shifting them onto the evolutionary tracks of rapidly evolving higher-mass central stars, and decreasing the estimated PN ages.}. In brief, the properties of the nebulae derived from the present spectroscopic observations are very similar to the 16 PNe studied earlier (Paper I).  As we shall see, all 18 PNe form a homogeneous group with similar masses, ages, and metallicities.  e shall see, all 18 PNe form a homogeneous group with similar masses, ages, and metallicities.  

Hereafter we refer to the ``oxygen abundance'' using the convention 12+log(O/H), where O/H is the result of the {\it icf} analysis.  The average abundance for the sample of 18 PNe is $8.62\pm0.14$.  The corresponding value for the Sun is 8.69 \citep{A09} and for the Orion Nebula it is 8.73 \citep{E04}.  Thus the ensemble-average O/H ratio is solar (within the uncertainties).  Fig.~1b displays the oxygen abundance gradient; no significant trend is obvious.  The slope of a simple linear fit is -0.005 dex kpc$^{-1}$ with a regression coefficient, {\it R}, of 0.41 and a standard deviation of the residuals $\pm 0.14$ (that is, fitting a linear slope does not decrease the scatter of the points from the trend line).  

In Paper~I, M1074 (denoted ``PN10'') was included in the data tables but omitted from the gradient fit because [O~II] $\lambda$3727 was not detected, yielding only an upper limit for O$^+$/H$^+$. In reviewing the data for M1074 we found that the upper limit of O$^+$/O is a tiny fraction of O$^+$+O$^{++}$, so the omission of O$^+$ in its O/H ratio is insignificant. Accordingly we have included it here for completeness. Had M1074 been included in Paper~I it would have decreased the slope of the gradient from -0.011 to -0.007 dex kpc$^{-1}$. 

\subsection{Comparison with overlapping spectroscopic surveys}  

Recently \citet{SC12} (``SC12") obtained spectra using the ``Hectospec'' fiber spectrograph of the MMT near Tucson for 713 heterogeneously selected H~II regions and PNe  in M31 and cross-referenced their targets to those of M06.  The way in which their sample of PNe was chosen precludes an unbiased PNLF from their measurements. SC12 derived O/H values for all 68 of the 459 of the PNe for which the [O~III] $\lambda$4363 flux was derived.  We compute that their average oxygen abundance is $8.41 \pm 0.25$.  This scatter is almost twice as large as we derive for our measurements. The gradient of their O/H abundances is +0.00386 dex kpc$^{-1}$ with a regression coefficient = 0.015; that is, consistent with a scatter diagram\footnote{There are Merrett numbers for 101 of their 253 H~II regions, suggesting that the M06 studies of the PNLF are contaminated by small numbers of tiny H~II regions in the inner disk.}. 

Both M174 and M2496 were observed by SC12 who derived a value of O/H for M174 that is larger than ours by 0.1 dex (well within the common uncertainties); they did not observe [O~II] $\lambda$3727 in M2496 and therefore did not determine its O/H. A comparison between our fluxes and theirs shows general agreement except for: (1) the fluxes of the relatively weak [O~III] $\lambda$4363 line of M174, for which the disagreement is 50\% (more than the sum of the errors\footnote{We suspect that confusion of this line with the $\lambda$4358 sky line of Hg~I is to blame when observing at moderate dispersion using a fiber spectrograph.  This sky line is faint at the Roque de los Muchachos Observatory.}); and (2) the fluxes of the bright H$\alpha$ and [O~III] $\lambda$$\lambda$4959,5007 lines for which we sometimes find fluxes 10\% larger than theirs.  We also find that the ratio of two bright lines, H$\alpha$ and H$\beta$, is consistently different by up to 10\% for five objects in common.  Their H$\alpha$/H$\beta$ ratio is sometimes $<$ 2.8, below the limit for case-B recombination lines for conditions in PNe.  As a consequence our extinction corrections and final results do not generally agree well. 

We compared the O/H abundances of five PNe observed by SC12 in common with our samples.  Ratios of O/H for individual PNe derived by SC12 and our group for individual PNe range from 0.4 to 1.8.  We have not attempted to resolve the systematic differences of our abundance results which are comprised a combination of uncertainties or differences in measured line fluxes, reddening corrections, derived temperatures, and the disparate methods used to extract ionic and total abundances from the data. 

\subsection{Additional Results}  The deep GTC spectra also yielded some unexpected detections of broad C~III and C~IV  lines in the stellar spectrum of M174 and M2496.  These are lines  commonly associated with the winds of WC stars; we present these results in Appendix B.

It is worth adding that in 2012 October we obtained spectra of three additional PNe in M31: M50, M2507 and M2549, which were observed between two and three hours each on the same night using the DIS spectrograph at Apache Point Observatory in generally clear weather but erratic seeing.  M50 lies just inside the inner range of our previous sample of PNe (R$_g$ $\approx$16 kpc).  The other two lie well beyond it (R$_g \approx 100$ kpc). Our goal was to see whether the general spectroscopic properties of these PNe were readily distinguishable from those of M174, M2496 and the other 16 PNe of KL12.  The innermost of the three PNe, M50, is sufficiently bright that we could derive an oxygen abundance of 8.7 $\pm$ 0.2 dex, in good agreement with other PNe at the same R$_g$.  The brighter lines of the other two objects, M2507 and M2549, are very similar to those of the other 18 PNe.  The [O~III] $\lambda$4363 lines  were detected but faint.  The measured [O~III] $\lambda$4363/$\lambda$5007 ratios support only a lower limit on their O/H abundances: $\geq$ 8.2 and 8.6, respectively.  In brief, the limits of O/H for M2507 and M2549 indicate no clear sign of any abrupt change of the O/H abundance ratio for PNe inside R$_g$ = 16 kpc or beyond $\approx$60 kpc. The results of the new APO data are preliminary.  We have been granted time to obtain deeper spectra of these and other PNe in M31using the 10.4-m GTC and 3.5-m APO telescopes in the autumn of 2013.

\section{Discussion}

Our primary goal in this section is to find a framework for interpreting the basic observables in this paper: (1) the uniformly solar-like O/H (and other low-mass $\alpha$ elements) in 18 bright PNe; (2) the general kinematics of the sample that much more closely resemble the rotation pattern of the classical disk of M31 than its halo or bulge, and (3) the uniqueness of the high O/H and flat gradient in our M31 ensemble of PNe relative to PNe in other galaxies.   Our first task is to integrate the present results into complementary information about the outer regions of M31.  We focus on studies of stellar populations since no significant reservoir of cold gas, dust, or ionized gas is found (\citealt{T04, I05, Mon09, Az11}).  Nor is there direct evidence of ongoing star formation in the disk of M31 beyond R$_g$$\approx$ 20 kpc (Choi et al. 2002, Cour11).  

\subsection{Population Membership}
As noted earlier, M174 and M2496 lie along M31's minor axis a degree (projected distance 14 kpc) from its center where their locations suggest that they might be members of M31's bulge or halo (Brown et al. 2008, Fig.~1) rather than an extended disk in which we have argued  the other 16 PNe are found (Paper~I).   By way of review note that the innermost dozen PNe in this study between $\sim 20 < $R$_g <  30$ kpc lie along an elliptical arc that has the same shape and orientation as the outer isophotes of the disk itself. Indeed, as Fig.~1a shows, the luminous PNe inside the 40-kpc ellipse have disk-like angular distribution. That strongly suggests  a direct association of these PNe with the disk.  Moreover, taken as a group, the set of 16 PNe studied initially shares the same kinematic patterns (and statistical deviations) as nearby disk stars (\citealt{Ib05}, Fig.~5 and \citealt{Ch06}, Fig.~1b) and scattered clouds of H~I \citep{T04}.   Such large scatter is characteristic of the thick disk of M31 \citep{Col11} and the disks of spiral galaxies that have been disrupted by impacts with smaller galaxies \citep{Sal09}. This all but eliminates the hypothesis that most of all of the sixteen PNe are or have been members of the metal-poor and kinematically inhomogeneous halo. 

Thus we now explore whether M174 and M2496 are  members of the same dynamical group as the other luminous PNe within the R$_g$ = 40 kpc ellipse of Fig.~1a.  Obviously their locations reveal little about their membership since M174 and M2496 were selected to lie along the minor axis.  The heliocentric Doppler motions of M174 and M2496 are -209 and -324 km s$^{-1}$, respectively.  These differ somewhat from those expected of ideal disk members in circular orbits ($\sim 300$ km s$^{-1}$)---but not by more than the other 16 PNe.  Although the spatial and kinematic evidence is inconclusive, the uniformly unusual metallicities of all 18 PNe allow us to presume that M174, M2496, and the remaining sixteen PNe are a physical group that might share a common history.  Thus we place both PNe in the extended disk of M31 (section 1) at deprojected distances R$_g > 55$ kpc from its center (Table~1).  

To further explore possible PN--stellar-population membership we exploit the abundance studies of RBG stars in the same general vicinity as the ensemble of PNe by Chapman et al. (2006, ``CH06").  CH06 obtained deep Keck spectra of nearly 10,000 RGB stars in 54 scattered fields, most all of them located outside the 20-kpc ellipse shown in Fig.~1a.  The [Fe/H] results that they found in each field are based on high-quality spectra of the Ca~II triplet lines in at least ten stars per field.   CH06 could not estimate the ages of their target stars from their spectra.

CH06 divided their sample into two groups: those with distinctly disk-like kinematics and those with spheroidal kinematics.  Most of their observed fields show [Fe/H] $< -0.7$.  This result is expected for the spheroidal stellar population \citep{K06}.  Of more interest here is the disk component.  In Fig.~9 of CH06 it can be seen that all five of the metal-rich disk fields ($0 <$ [Fe/H] $< -0.5$) are in the outer zone between the 20-kpc and 40-kpc ellipses of Fig.~1a. (That is, the [Fe/H] gradient has a {\it positive} slope.)  Moreover, the metal-rich fields each lie to the northwest of the 20-kpc ellipse; that is, in the immediate vicinity of 13 of the PNe studied in Paper~I in which the O/H abundances are nearly solar. This shows a high spatial correlation of high-metallicity RGB stars and most of our oxygen-rich PNe. 

The potentially high metallicity of disk stars at about the same R$_g$ is confirmed by the analysis of very deep color-magnitude diagrams (``CMD'') obtained by \citet{BS06}  with the Advanced Camera for Surveys (``ACS'') on the Hubble Space Telescope (``HST'').  They analyzed spheroidal and disk-dominated fields at R$_g \approx$ 30 kpc.  After subtracting the spheroidal contamination in the disk field, BS06  found that the relatively metal-rich disk population is characterized by [Fe/H] $\approx$ +0.3 (and its age is estimated at 6 GY). The disk field is located along M31's major axis about 28 kpc to the northeast of the nucleus where we are planning future observations of luminous PNe.

Similarly, \citet{Ri08} and \citet{Ri09} constructed CMDs from deep ACS images in several fields in M31 far beyond R$_{25}$.  Their fields lie along the edge of the extended disk.  This outer stellar disk is characterized by an exponential scale length of 14 kpc, a semi-major axis of 55 kpc (similar to the deprojected radii of M174 and M2496), and the general colors of a metal-rich stellar population \citep{I05}.  The CMDs at the edge of this extended disk constructed by Richardson et al. show that the metallicity of stars at the outer edge of has an irregular distribution at that projected radius.  Values of [Fe/H] derived by \citet{Ri09} in these and more distant fields show significant variations from $\approx$-0.54 to -1.03 with only a small correlation of the metallicity with the presence of faint stellar streams and related features.  

Recently Bernard et al. (2012, ``BFB12'')  published deep ACS CMD studies of the stellar populations in two fields on either side of M31's H~I warp.  They concluded that a relatively young ($\approx$2 GY) stellar population is present in the warp field that is characterized by solar [Fe/H].   Interestingly, another CMD of a field located inside M31's warp radius showed no trace of a young and high-metallicity counterpart. 

\subsection{$\alpha$-element abundances of PNe in other spiral galaxies}  

Our second task is to investigate whether the O/H trends found in our M31 ensemble of PNe is typical of other galaxies.  The ensemble-average value of O/H for the PNe in M31 are essentially the same as the corresponding values for PN in disk galaxies of comparable luminosity; that is, the value of $<$12+log(O/H$>$ for the PNe in the outer disk of M31 is in  agreement with or slightly higher than those of the inner disks of the Milky Way (8.61;  \citealt{KH12} [``KH12"]),  M81 (8.59; Stang10), and NGC~300 (8.57; Stas13).  All of these O/H results are well in excess of those in less massive galaxies, including M33 (8.29; Mag09), the LMC (8.27; KH12), the SMC (8.04; KH12) and M32 ($\approx$8.3, \citealt{RM08}, with considerable dispersion in the data, as estimated from their Fig.~4).

However, the O/H values of PNe in M31 differ from those in other systems in that M31 shows little if any sign of a radial gradient.  Mag09 and Stang10 found that the radial gradients in M33 and M81, respectively, are clearly established and in line with those of H~II regions in the same ranges of R$_g$.  These authors concluded that there is no evidence of recent oxygen enrichment in either galaxy.  PNe in NGC~300 also show a somewhat negative O/H gradient (Stas13), though its slope is flatter than that of nearby H~II regions.  From this Stas13 concluded that oxygen is currently undergoing enrichment in its disk.

From their analysis of N and O gradients of the PNe Stas13 conclude ``oxygen is affected by nucleosynthesis in the PN progenitors, by an amount which depends at least on the stellar rotation velocity and possibly other parameters''. In other words, PNe in NGC~300 are enhancing their oxygen, in contradiction to many common models of nucleosynthesis (next section) and the studies of M31, M81, and the Milky Way cited in \S 1\footnote{It is worth noting that the PNe in all of these galaxies as well as and those in M31 (Paper~I), the SMC and LMC (KH12), and the Milky Way (H10) --as well as the Sun and Orion -- share the same Ne/O ratios. This indicates that O and Ne are universally enriched at the same rates and, presumably, by the same processes in all of these galaxies, implying that any oxygen enrichment process operating within NGC~300 (and apparently only NGC~300) also proportionally enriches Ne.  See the discussion in section 7 of Stas13 for  more details.}.

\section{Interpretation}

The science goal of this section is to uncover---or at least to illuminate---the origins of the highly enriched PNe in this study.  There are some hard constraints: The anomalously high O/H ratio and flat radial gradient of all eighteen PNe in M31 suggests that an that a common explanation of their solar-like O/H abundances is necessary.  Moreover, we seek an explanation of the abundance pattern of PNe of our M31 ensemble that is not characteristic of other spiral galaxies whose radial O/H abundance gradient is steeper.  Thus we seek a scenario that connects our results to a particular history for M31's outer regions. In particular, we are looking for some sort of a ongoing mechanism that transports the metal-rich stars or ISM in the inner disk into the outer extended disk after nuclear burning processes had a chance to enrich the metals.  This may take the form of an event that deposited stars or gas with $\alpha$ elements with approximately solar abundances in the other disk.    

\subsection{Injection of enriched gas and stars in M31's outer disk} 
It is instructive to identify and eliminate some common enrichment scenarios that don't seem to apply to M31.  For example, we have just seen that not only the PNe, but also many nearby RGB stars share similar metallicities, so it seems unlikely that PNe in M31 have undergone self-enrichment of oxygen.  

In addition, the process of radial migration (\citealp{SB02, ScB09, ScB092, L11, Bird12}) has been suggested as a means to  transport stars and gas radially outward from inner metal-rich parts of the disk.  However, migration acts gradually and only where the gravitational perturbations of the arms are prominent.   This mechanism isn't likely to account for the flat O/H gradient or patchy groups of metal-rich stars found far beyond the inner disk.  Moreover, \citet{2010ApJ...712..858G} found that the migration hypothesis was inapplicable to the PNe in the disk galaxy NGC~300.  Finally, the process is incapable of producing the positive [Fe/H] gradient observed in the disk component of RGB stars studied by CH06.  Further, radial migration is likely to operate similarly in all comparable spirals (esp. M33 and M81 in which the spiral arms are more prominent than in M31), contrary to the differences of O/H gradients in other spirals noted in section 4.  

Local ISM enrichment by core-collapse supernovae is implausible and never observed in the outer parts of large spiral galaxies.  A gaseous transport process known as the ``fountain effect'' can redistribute enriched disk gas using the momentum of supernova blast waves as a transport mechanism.  However, \citet{SM13} concluded that fountains in the inner disk of M31 could only mix gas locally ($\approx$1 kpc).   

Metal-rich stars and gas can be transported outward by external mechanisms; to wit, galaxy-galaxy encounters.  The outer regions of M31 are marbled with stellar streams and other structures that are generally assumed to be the result of numerous past encounters with dwarf galaxies.  The ``splashes'' following their impacts can also disrupt the ambient ISM of the inner disk \citep{P10}.  However, random bombardments by dwarf galaxies only eject and disperse small amounts of ISM to high galactic latitudes---like stones thrown in a lake.    Even many encounters with dwarf galaxies are not likely to account for the high metal content of all of the PNe in the ensemble or the many metal-rich stars far from the inner disk.  

So we seek some sort of rare, major, recognizable and probably fairly recent event that led to a major disruption of the ambient metal-enhanced gas and the possibly the stars in the inner disk. We now review the evidence---of which there is a growing body---for an encounter between M31 and a gas-rich galaxy that triggered a starburst from which the ensemble of PNe in this paper evolved.  Our aim is to find a ``smoking gun'': an event identified by its age and degree of disruption of both galaxies involved in the encounter.  This can be inferred from a plausible model of the interaction combined with estimates of the ages of the PNe and the metal-rich stars.

\subsection{The enrichment of gas and stars in the outer disk of M31}  
\citet{2008MNRAS.386..461C} modeled the plausible consequences of a future collision of M31 and the Milky Way.  They stated: ``Eventually, after the merger has completed, the Sun is most likely to be scattered to the outer halo and reside at much larger radii ($> 30$ kpc)".  Similarly the formation of a starburst in M31 that could lead to the production of a generation of metal-rich group of PNe might have resulted from a comparable encounter with another large galaxy in the past. 

The obvious candidate is the spiral galaxy M33 presently located about 14$^{\circ}$ ($\approx 200$ kpc) southeast of M31.  McConnachie et al. (2009, ``Mc09'') obtained very deep tiled images of a 220 square degrees in an extended field containing both M31 and M33 as well as a three-color image overlay around M31.  They found a bridge of RGB stars  that includes a prominent stellar stream joining the two galaxies.  In an M31--M33 analogue of Cox \& Loeb's models, Mc09 ran a ``soft-particle hydrodynamical'' model (``SPH'') which demonstrated that an encounter of M31 and M33 about 2.6 GY ago (pericenter $\sim 53$ kpc) could account for the morphologies of the warp in the inner disk of M31 and some of its stellar streams\footnote{A very instructive simulation of the stellar dynamics during and after this encounter by J. Dubinski and L. Widrow is available at  www.youtube.com/watch?v=6eZm3LHlyrs; pericenter = 53 kpc at -2.6 GY.} (the SPH simulations do not track the ejected gas or the origins of starbursts).  The model of the interaction shows that stars from the disk can be disrupted to the outer regions of M31 that were discussed in \S 4.   \citet{BT04} and \citet{LF12} have found a corresponding bridge of H~I that spatially and kinematically connects M31 and M33.

The ages of stellar populations can be untangled from high-quality CMDs.  \citet{Ham10} interpreted existing CMDs from the literature and the morphologies of stellar streams outside of M31's extended disk to suggest similarly, that a merger with a 3:1  mass ratio at a pericenter of 25 kpc unfolded between 5.5 and 9 GY ago.  The predominant ages of metal-rich stars ([Fe/H] $\sim +0.3$), in the outer disk found by \citet{BS06}, 5 -- 8 GY, serve as evidence of a post-encounter starburst. 

 \citet{Ri08} argue that stars in the field EC1 on the northwest side of M31 (near many of the PNe in our sample) were stripped from the progenitor galaxy of the Giant Stream that extends 50\% of the way to M33.  Their relatively low values of [Fe/H] (-0.8) suggest that these stars migrated from an old stellar population the outer disk, possibly during an encounter.  
 
 More recently, BFB12 found that a high-quality CMD of a field at the outer edge of the disk in M33 (originally published by \citealt{Bar11}) contains a metal-rich, $\approx$2 GY-old stellar population---much the same as that found from their deep CMD in their warp field of M31.  These unusual findings are best explained by a common origin---the M31--M33 encounter modeled by Mc09---and establish that the starburst occurred from a highly metal-enriched ISM. 

The recent M31--M33 encounter scenario is also corroborated indirectly.  On the observational side, even though the H-band flux of M33 is 15 times fainter than that of M31 (NED), M33 has a current global rate of star formation that is 4.6 times higher \citep{K98}. This supports the notion that a massive starburst in M33 has been triggered by a recent global disruption.  Also, \citet{Ci09} found that the stellar population of M33 shows that the [Fe/H] gradient flattens abruptly beyond the spiral arms (R$_ g <$ 9 kpc).  This is an expected outcome of the numerical model of the M31--M33 encounter by Mc09. 

H~II regions of other disrupted disk galaxies have been found to show a pattern of flat O/H gradients.  \citet{W11} measured the O/H ratios of H~II regions throughout 13 galaxies from the ``Rogues H~I Survey.''  The target galaxies were selected for their unusually extended H~I disks and/or peculiar H~I features indicative of recent galaxy encounters \citep{Hib01}. Werk et al.  found that the H~II regions in the Rogues sample show emission-line spectra indicating higher-than-expected O/H at large R$_g$.  They suggested that the ionizing stars formed from metal-rich material that was transported to larger galactic radii by the encounter that disturbed the H~I out to the regions where the present H~II regions formed.  Although they employed somewhat unreliable ``strong-line'' methods for deriving the absolute value of O/H (\citealt{{B11}, {LS10}, {P06}}), the finding of a flat O/H gradient in their target galaxies is almost certainly a secure conclusion since it is based on a single methodology.  However, as shown by \citet{SC12}, the application of various strong-line methods to the same set of data yields disparate results for O/H.  Accordingly, and in light of the considerable scatter in the O/H values for H~II regions compiled in Fig.~1b for M31, we have not performed a comparative analysis of the O/H abundance measurements between PNe and H~II regions.

\subsection{The formation of the ensemble of oxygen-rich PNe in M31} 
The next question is whether the ages of PNe connect their progenitor stars to either of the encounter events $\sim2 $ or $\sim 6$ GY in the past.   The method for estimating the ages of the progenitors was described in Paper~I.  In summary, it relies on stellar evolution models that predict the evolving ultraviolet luminosities and stellar temperatures of post AGB stars that shed and later ionize the nebular gas.  Models by \citet{RM97}, \citet{Mar01}, \citet{Mar04}, \citet{Ca08}, and \citet{K10} (and others) predict the evolving AGB mass yields as a function of the initial mass of the progenitor star.  Other models such as those of \citet{Sc92}, \citet{Mar04}, \citet{S05}, \citet{S07}, and \citet{Men08} predict the evolving properties and related observables of the nebula.  These observables can used to locate the central stars of PNe on an H--R diagram and, hence, to determine their initial masses and ages from evolutionary tracks.  

Following this procedure we find that the characteristic initial masses of the central stars are on the order of 1.5 M$_\odot$ and their corresponding evolutionary ages are $\sim 2$ GY. This is the same age derived by other authors who selected their targets from the PNe with highest L$_{[OIII]}$, including Mag09, Stang10, and Stas13.  The ages of these PNe are sufficiently short that their $\alpha$-element abundances must be very close to those of the current ISM.

Therefore, taken at face value, associating the ensemble of PNe with the M31--M33 encounter of 2.6 GY ago (Mc09) and post-encounter metal-rich starburst of 2 GY ago (BFB12) is straightforward.  Although the evidence, such as it is, is largely circumstantial, the puzzle pieces fit without mishap.  If the $\sim2$-GY ages of the PNe are appropriately estimated then the ensemble of PNe cannot be attributed to any earlier starburst in M31.

The veracity of this or any other PN age estimate method can be challenged owing to the universality of the bright-end cutoff of the PNLF.  Specifically, the form of the PNLF has been found to be independent of the size and stellar content of the underlying galaxy, including some early-type galaxies in which star formation has been dormant for cosmic time scales.  (See \citealt{C05} and \citealt{C10} for a much deeper discussion of the challenges this poses to current understanding of stellar evolution.)  In other words, PNe selected for their high [OIII] luminosities, such as those in this study, have indeterminate ages.  While their O/H enrichment might be explainable by metal-enriched gas transported to the outer disk by a past encounter, their formation cannot be directly tied to any particular event.  

Despite their common abundances and kinematics, this age indeterminacy may well apply to PNe in M31. M06 found that the radial distribution of PNe follows the distribution of R-band light out to a radius of 2$^{\circ}$ along M31's major axis (6 disk scale lengths = 0.9 R$_{25}$). This and the similarity of R, I, J, and K surface brightnesses along M31's major axis (Cour11) suggest that PNe are associated with a very old stellar population of M31.  M06 also found that the shape of the PNLF was invariant with deprojected disk radius, R$_g$, within their densely sampled zones along the entire major axis of the inner disk of M31.  While the present sample of PNe mostly lie beyond this radius, and while their numbers are too small to check the shape of the PNLF, we have no reason to argue that the PNLF of the outer disk of M31 is anything but ordinary.

Until the enigma of the universal PNLF is resolved---that is, why otherwise successful theories of AGB and post-AGB evolution cannot be used to predict the progenitor mass---we can only conclude that all eighteen O-rich PNe in M31 seem to have had a common origin, and one that is most likely the result of a past starburst event within a metal-rich environment.

\section{Acknowledgments}
The remarks and suggestions of the referee of this paper, Robin Ciardullo, have led to material improvements in its science content.  We deeply appreciate his considerable efforts and always-helpful critical comments. 

The results of this paper are based on observations made with (1) the Gran Telescopio Canarias (GTC), installed at the Spanish Observatorio del Roque de los Muchachos of the Instituto de Astrof\'isica de Canarias, in the island of La Palma and (2) The 3.5-m telescope at Apache Point Observatory in Sunspot NM.  B.B., K.B.K., and R.B.C.H. are grateful to our institutions and to the NSF for support under grants AST-0806490, AST- 0808201, AST-0806577, respectively.  R.L.M.C acknowledges support from the Spanish AYA2007-66804 and AYA2012-35330 grants. This research has made use of NASA's Astrophysics Data System.

\section{Appendix A.  Tabulations of Observations and Results}
As stated in \S3 these data have been analyzed using the same techniques as those in KL12, so that the results are directly comparable to the other PNe in Fig.~1a. 
 %The first step in the analysis is to generate a table of line intensities that have been corrected for interstellar reddening and for contamination of the hydrogen Balmer lines by coincident recombination lines of He$^{++}$. 
Table~A1 contains the emission-line measurements. Column entries are as follows: the first column lists the ion and wavelength designation of each line; f($\lambda$) gives the value of the reddening function, normalized to c(H$\beta$) = 0; F($\lambda$) is the measured flux and estimated error, relative to H$\beta$ = 100; and I($\lambda$) gives the reddening-corrected intensity and estimated error, also relative to H$\beta$ = 100. Interstellar reddening corrections are based on \citet{SM79}.  At the bottom of each column, for each nebula we list the logarithmic reddening parameter, c(H$\beta$), the theoretical H$\alpha$/H$\beta$ ratio appropriate for the nebular temperature and density, and the log of the total observed H$\beta$ flux through the spectrograph slit.

Table~A2 contains the ionic abundances for each of the PNe.  It also shows the temperatures adopted from Table~A3 that were used to calculate them.  Asterisks denote values that were used in the brightness-weighted mean values in lines below them.  The derived ionization correction factor ({\it icf}), calculated as described in \citet{KH01} is shown for each element.  This shows the factor applied to the sum of the measured ionic abundances used to correct for unobserved ionization states of each atom.  Table~A3 lists all the available diagnostics and total derived abundances for each PN.

The physical descriptors of M174 and 2496, temperature T$_e$ and density N$_e$, as well as their abundance ratios fall into line with the other 16 PNe that were studied in KL12.  All 18 of the PNe form a homogeneous group with very similar abundance ratios. 

The stellar temperature, luminosity, initial and final mass, and evolutionary age were estimated from Cloudy model fits to the data using procedures described by KL12, \S 3.3.1.  The results are shown in Table~A4.  We adopted log g = 6.5 and truncated the density distribution in order to match the observed flux of [O~II] $\lambda$3727.  The relatively large stellar mass of M2496 is consistent with its bright [O~III] magnitude, m$_{5007}$ (\S2).  Barring errors in the fitted model, the age of this PN is surprisingly small.  We find that the locations of M174 and M2496 fall slightly to the right of the other 16 PNe on an H-R diagram; see Fig.~3 of KL12.

%The important point is that these measurements place the central stars near the evolutionary tracks for final masses of 0.6 M$_{\odot}$ and independently confirm the initial masses of the central stars near 2 M$_{\odot}$, as claimed in \S1.

\section{Appendix B.  Carbon Lines in the Spectra of the Central Stars}
The spectra of both M174 and M2496 show broad emission features that can be attributed to central stars of Wolf-Rayet (WR) type (Fig.~B1). In both objects the strongest feature is C~IV $\lambda$5805, with full width at half maximum (FWHM) $\sim$35\AA~. Also visible in both is C~III $\lambda$4649 which, in PN174, has a red shoulder that is likely due to He~II $\lambda$4686. No signs of other broad emission features (in particular, C~III $\lambda$5696 or oxygen lines) are seen in our spectra: their upper limit is estimated to be 10\% of the C~IV $\lambda$5805 flux. The line ratios and FWHM, shown in Table~B1, indicate that both stars are type [WC4] according to the scheme of \citet{A03}.  This type corresponds to stellar temperature between 50,000 K and 90,000 K, consistent with the modeled values around 70,000 K. 

PN2496, only 0.2 mag from the bright cutoff of the M31 [O~III] PNLF, is one of the brightest PNe in M31 (M06).  PN174 is also bright in [O~III], 0.7 mag from the  cutoff.  In this respect, finding that their central stars are of WR type comes as a surprise. According to the most recent attempts to model the PNLF \citep{S07}, PNe with He burning central stars are not expected at its bright end. This conclusion is based on the limited information available for He burners [e.g. \citet{VW94}], according to which these central stars are less luminous than hydrogen-burning models when they reach high temperatures, and evolve more slowly, thereby favoring the development of optically thin (i.e. less luminous) nebulae.  Our detection of WR features indicates that, on the contrary, He burners also manage to have a combined stellar and nebular evolution so as to reach the large [O~III] luminosities near the PNLF cutoff.

It is unlikely that a very late thermal pulse (VLTP) is at the origin of the WR nature of the central stars of PN174 and PN2496, as the nebulae have high density, i.e., are presumably young, and are hydrogen-rich.  A final thermal pulse when the progenitor is about to leave the AGB (AFTP), or a late thermal pulse (LTP) early in the post-AGB track, is favored instead.

These PNe are not the first in M31 in which WC lines have been observed.  \citet{JC99} found typical WC6 emission lines in their PN FJCHP 57.  Although this PN is located near the edge of the M31's disk, its discordant Doppler shift suggests it is a member of the halo.  \citet{Mag09} also found that PN039 in M33 shows WC stellar lines.
 
Note that \citet{G09} found that PNe with WR central stars are among the brightest ones in the Galactic bulge, and so are some WR PNe in the Large Magellanic Cloud [see e.g., \citet{M98}]. Together with PN174 and PN2496 in M31, the occurrence of PNe with WR-type central stars in the most luminous PNe suggests that the general evolution of He burners and their nebulae are not yet adequately understood.

\begin{figure}
\epsscale{1}
\figurenum{1}
\plotone{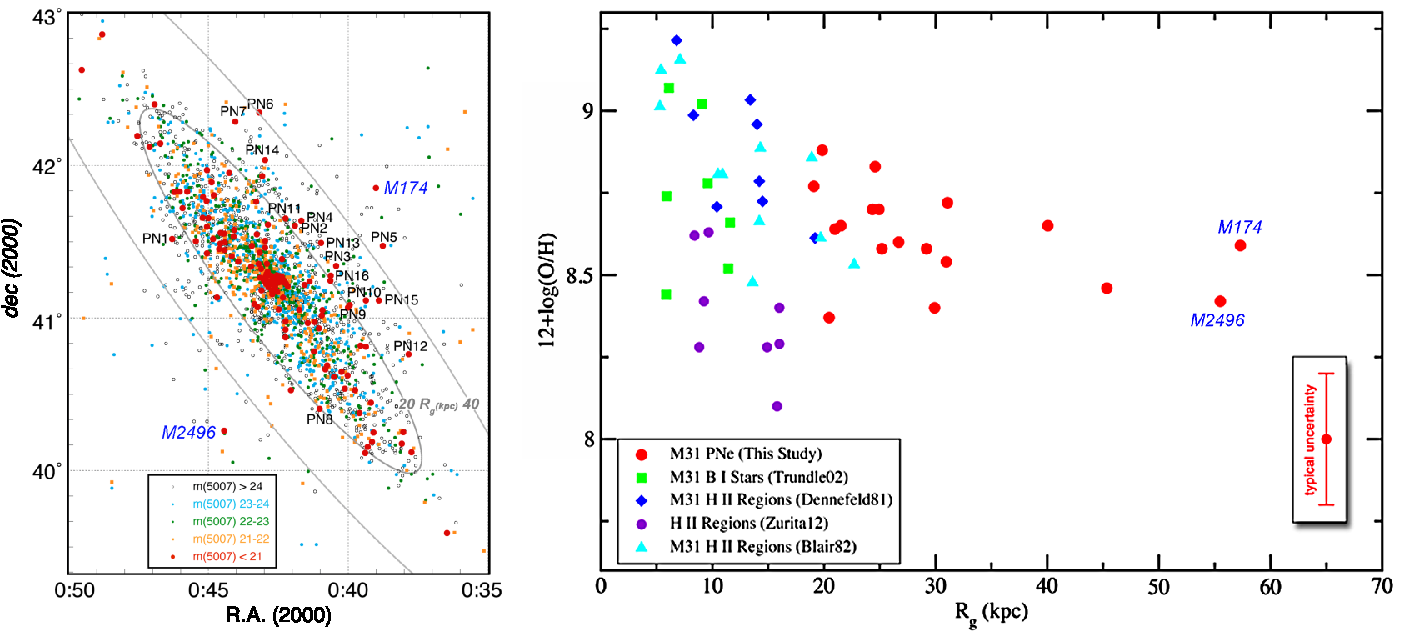}
\caption{Left panel (1a): The locations of M31 PNe with the locations of the GTC targets M174 and M2496 shown with blue labels and the PNe studied by KL12 noted with black labels.  The figure is adapted from their Fig.~1.  Right panel (1b): Values of 12+log(O/H) plotted against deprojected (``rectified'') galactocentric distance R$_g$ for various objects in M31.  The red dots are the 18 PN targets of this and KL12.  The PNe observed using the GTC are the two rightmost red dots labelled in blue.  Values of 12+log(O/H) and R$_g$ for H~II regions and stars are taken from the original data sources.   The color version of this figure appears in the on-line manuscript.}
\end{figure}

%\begin{figure}
%\epsscale{0.7}
%\figurenum{2}
%\plotone{Fig2kk.eps}
%\caption{Values of 12+log(O/H) plotted against deprojected (``rectified'') galactocentric distance R$_g$ for various objects in M31.  The red dots are the 18 PN targets of this and KL12.  The PNe observed using the GTC are the two rightmost red dots labelled in blue.  Values of 12+log(O/H) and R$_g$ for H~II regions and stars are taken from the original data sources.  The color version of this figure appears in the on-line manuscript.}
%\end{figure}

\begin{figure}
\epsscale{0.5}
\figurenum{B1}
\plotone{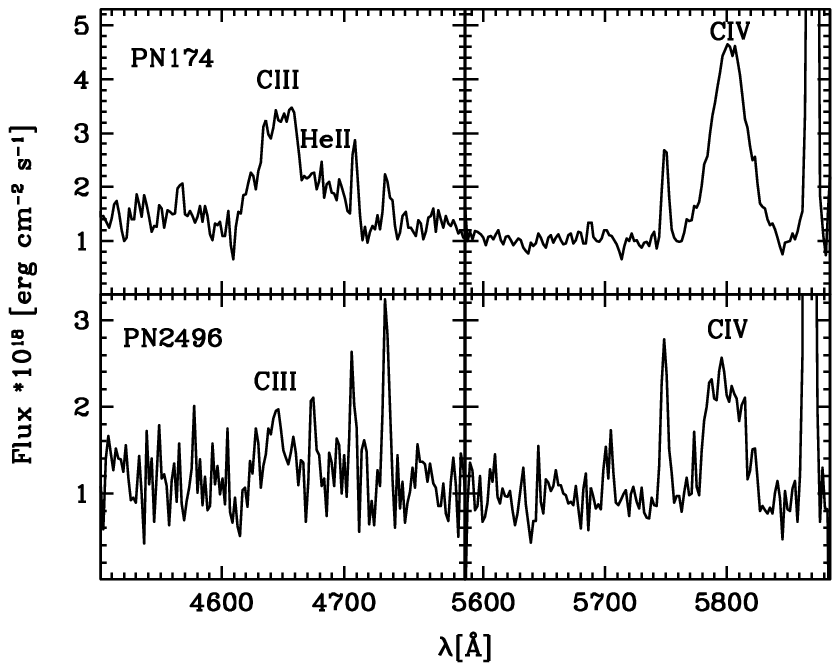}
\caption{Close-up of the spectra around the WR broad emission features of the central stars.}
\end{figure}

\clearpage

\begin{deluxetable}{lcccccc}
\tabletypesize{\scriptsize}
%\rotate
\tablecolumns{7}
\tablewidth{0pc}
\tablenum{1}
\tablecaption{Basic and Computed Properties of the Target PNe}
\tablehead{
\colhead{Merrett \#} &
\colhead{RA$_{2000}$} &
\colhead{DEC$_{2000}$}&
\colhead{m$_{5007}$\tablenotemark{a}} &
\colhead{V$_{helio}$\tablenotemark{b}} &
\colhead{Angular offset\tablenotemark{c}} &
\colhead{R$_g$ (kpc)\tablenotemark{d}} 
}

\startdata
M174	&	0:39:01.0	&	 41:51:11.1 &	20.89 &	-209.0	&	0.91$^{\circ}$ &	57.6 \\
M2496	&	 0:44:26.0 	& 40:15:45.6 &	20.42 &	-324.0	&	1.06$^{\circ}$ &	55.5 \\
\enddata
\tablenotetext{a}{ m$_{5007}$ =  -2.5logF$_{5007}$ - 13.74}
\tablenotetext{b}{ taken from \citet{m06}}
\tablenotetext{c}{ measured from M31's center}
\tablenotetext{d}{ assumes D$_{M31}$=770 kpc \citep{FM90} and disk inclination = 77.7$^{\circ}$}
\end{deluxetable}

\clearpage
\begin{deluxetable}{lccccc}
\tabletypesize{\normalsize}
\setlength{\tabcolsep}{0.07in}
\tablecolumns{6}
\tablewidth{0in}
\tablenum{A1}
\tablecaption{Fluxes and Intensities}
\tablehead{
\colhead{} & \colhead{} &
\multicolumn{2}{c}{PN174} &
\multicolumn{2}{c}{PN2496} \\
\cline{3-4} \cline{5-6}  \\
\colhead{Line} &
\colhead{f($\lambda$)} &
\colhead{F($\lambda$)} &
\colhead{I($\lambda$)} &
\colhead{F($\lambda$)} &
\colhead{I($\lambda$)}}
\startdata
\[[O II] $\lambda$3727 & 0.292 & 47.5 & 49.4$\pm$11.01 & 7.65 & 8.41$\pm$1.88 \\
He II + H11 $\lambda$3770 & 0.280 & 3.63 & 3.77$\pm$0.83 & \nodata & \nodata \\
He II + H10 $\lambda$3797 & 0.272 & 4.04 & 4.19$\pm$0.91 & 6.08 & 6.64$\pm$1.44 \\
He II + H9 $\lambda$3835 & 0.262 & 10.9 & 11.3$\pm$2.40 & \nodata & \nodata \\
\[[Ne III] $\lambda$3869 & 0.252 & 83.4 & 86.3$\pm$18.13 & 104 & 113$\pm$24 \\
He I + H8 $\lambda$3889 & 0.247 & 20.1 & 20.8$\pm$4.33 & 14.2 & 15.3$\pm$3.19 \\
\[[Ne III] $\lambda$3968 & 0.225 & 23.8\tablenotemark{a} & 24.6$\pm$8.15\tablenotemark{a} & 36.1\tablenotemark{a} & 38.9$\pm$11.05\tablenotemark{a} \\
H$\epsilon$ $\lambda$3970 & 0.224 & 15.5\tablenotemark{a} & 16.0\tablenotemark{a} & 14.9\tablenotemark{a} & 16.1\tablenotemark{a} \\
He I + He II $\lambda$4026 & 0.209 & 2.67 & 2.74$\pm$0.54 & 2.99 & 3.19$\pm$0.63 \\
\[[S II] $\lambda$4071 & 0.196 & 3.05 & 3.13$\pm$0.60 & 4.99 & 5.31$\pm$1.02 \\
H$\delta$ $\lambda$4101 & 0.188 & 24.3 & 24.9$\pm$4.74 & 27.7 & 29.4$\pm$5.59 \\
C II $\lambda$4267 & 0.144 & 1.74 & 1.77$\pm$0.31 & \nodata & \nodata \\
H$\gamma$ $\lambda$4340 & 0.124 & 40.1 & 40.8$\pm$6.99 & 44.4 & 46.2$\pm$7.92 \\
\[[O III] $\lambda$4363 & 0.118 & 7.78 & 7.90$\pm$1.34 & 15.3 & 15.9$\pm$2.71 \\
He I $\lambda$4472 & 0.090 & 5.21 & 5.27$\pm$0.86 & 5.37 & 5.530.90 \\
C III + O II $\lambda$4650 & 0.045 & 10.1 & 10.2$\pm$1.54 & \nodata & \nodata \\
\[[Fe III] $\lambda$4658 & 0.043 & 6.41 & 6.45$\pm$0.97 & \nodata & \nodata \\
He I + \[[Ar IV] $\lambda$4711 & 0.030 & 0.932 & 0.936$\pm$0.138 & 0.653 & 0.660$\pm$0.098 \\
\[[Ar IV] $\lambda$4740 & 0.023 & 0.862 & 0.864$\pm$0.126 & 1.66 & 1.67$\pm$0.24 \\
H$\beta$ $\lambda$4861 & 0.000 & 100 & 100 & 100 & 100 \\
He I $\lambda$4922 & -0.021 & 1.25 & 1.24$\pm$0.17 & 1.31 & 1.30$\pm$0.18 \\
\[[O III] $\lambda$4959 & -0.030 & 361 & 359$\pm$49 & 436 & 432$\pm$59 \\
\[[O III] $\lambda$5007 & -0.042 & 1078 & 1072$\pm$144 & 1307 & 1290$\pm$173 \\
\[[Cl III] $\lambda$5518 & -0.157 & 0.312 & 0.305$\pm$0.037 & \nodata & \nodata \\
\[[Cl III] $\lambda$5538 & -0.161 & 0.735 & 0.720$\pm$0.088 & \nodata & \nodata \\
\[[N II] $\lambda$5755 & -0.207 & 1.09 & 1.06$\pm$0.13 & \nodata & \nodata \\
C IV $\lambda$5806 & -0.217 & 12.2 & 11.9$\pm$1.47 & 6.62 & 6.17$\pm$0.76 \\
He I $\lambda$5876 & -0.231 & 18.0 & 17.5$\pm$2.17 & 18.6 & 17.2$\pm$2.14 \\
\[[O I] $\lambda$6300 & -0.313 & 7.50 & 7.19$\pm$0.96 & 7.63 & 6.89$\pm$0.92 \\
\[[S III] + He II $\lambda$6312 & -0.315 & 2.11 & 2.02$\pm$0.27 & 2.97 & 2.69$\pm$0.36 \\
\[[O I] $\lambda$6364 & -0.325 & 2.32 & 2.22$\pm$0.30 & 2.52 & 2.27$\pm$0.31 \\
\[[N II] $\lambda$6548 & -0.358 & 13.1 & 12.5$\pm$1.77 & 6.26& 5.58$\pm$0.79 \\
H$\alpha$ $\lambda$6563 & -0.360 & 298 & 284$\pm$1 & 315 & 281$\pm$1 \\
\[[N II] $\lambda$6584 & -0.364 & 40.2 & 38.3$\pm$5.47 & 19.8 & 17.6$\pm$2.52 \\
He I $\lambda$6678 & -0.380 & 4.11 & 3.90$\pm$0.57 & 4.35 & 3.85$\pm$0.56 \\
\[[S II] $\lambda$6716 & -0.387 & 1.49 & 1.42$\pm$0.21 & 0.981 & 0.866$\pm$0.128 \\
\[[S II] $\lambda$6731 & -0.389 & 3.13 & 2.97$\pm$0.44 & 2.17 & 1.92$\pm$0.28 \\
He I $\lambda$7065 & -0.443 & 9.90 & 9.33$\pm$1.50 & 13.2 & 11.4$\pm$1.84 \\
\[[Ar III] $\lambda$7136 & -0.453 & 14.2 & 13.4$\pm$2.19 & \nodata & \nodata \\
He I $\lambda$7281 & -0.475 & \nodata & \nodata & 2.17 & 1.86$\pm$0.32 \\
\[[O II] $\lambda$7324 & -0.481 & 14.8 & 13.9$\pm$2.38 & 18.4 & 15.7$\pm$2.70 \\
\[[Ni II] $\lambda$7378 & -0.489 & 3.58 & 3.35$\pm$0.58 & \nodata & \nodata \\
\[[Ar III] $\lambda$7751 & -0.539 & 2.85 & 2.65$\pm$0.50 & 2.53 & 2.13$\pm$0.40 \\
c & & & 0.06 & &0.14 \\
H$\alpha$/H$\beta$ & & & 2.84 & &2.81 \\
log F$_{H\beta}$\tablenotemark{b} & & -14.99 & &-15.08 \\
\enddata
\tablenotetext{a}{Deblended.}
\tablenotetext{b}{ergs\ cm$^{-2}$ s$^{-1}$ in our extracted spectra}
\end{deluxetable}

\clearpage

\begin{deluxetable}{lcclccl}
\tabletypesize{\small}
\setlength{\tabcolsep}{0.07in}
\tablecolumns{6}
\tablewidth{0in}
\tablenum{A2}
\tablecaption{Ionic Abundances}
\tablehead{
\colhead{} &
\multicolumn{3}{c}{PN174} &
\multicolumn{3}{c}{PN2496} \\
\cline{2-4} \cline{5-7}  \\
\colhead{Ion} &
\colhead{T$_{\mathrm{used}}$} &
\colhead{Abundance} &
\colhead{} &
\colhead{T$_{\mathrm{used}}$} &
\colhead{Abundance} &
%\colhead{Notes}
}
\startdata
He$^{+}$ & [O III] & 0.116$\pm$0.015 &  & [O III] & 0.104$\pm$0.014& \\
icf(He) &  & 1.00 &  &  & 1.00 & \\
\\
O$^{0}$(6300) & [N II] & $^{*}$1.04$\pm$0.81(-5) &  & [N II] & $^{*}$6.65$\pm$1.35(-6)  \\
O$^{0}$(6363) & [N II] & $^{*}$1.01$\pm$0.79(-5) &  & [N II] & $^{*}$6.84$\pm$1.41(-6)  \\
O$^{0}$ & wm & 1.03$\pm$0.80(-5) &  & wm & 6.69$\pm$1.30(-6)  \\
O$^{+}$(3727) & [N II] & $^{*}$4.11$\pm$7.63(-5) &  & [N II] & $^{*}$5.41$\pm$1.19(-6)  \\
O$^{+}$(7325) & [N II] & $^{*}$4.37$\pm$4.52(-5) &  & [N II] & $^{*}$2.59$\pm$0.73(-5)  \\
O$^{+}$ & wm & 4.17$\pm$6.83(-5) &  & wm & 1.99$\pm$0.53(-5)  \\
O$^{+2}$(5007) & [O III] & $^{*}$3.53$\pm$0.95(-4) &  & [O III] & $^{*}$2.48$\pm$0.64(-4)  \\
O$^{+2}$(4959) & [O III] & $^{*}$3.42$\pm$0.77(-4) &  & [O III] & $^{*}$2.40$\pm$0.50(-4)  \\
O$^{+2}$(4363) & [O III] & $^{*}$3.53$\pm$0.95(-4) &  & [O III] & $^{*}$2.48$\pm$0.64(-4)  \\
O$^{+2}$ & wm & 3.50$\pm$0.88(-4) &  & wm & 2.46$\pm$0.59(-4)  \\
icf(O) &  & 1.00 &  &  & 1.00 \\
\\
Ar$^{+2}$(7135) & [O III] & $^{*}$1.18$\pm$0.27(-6) &  & [O III] & \nodata  \\
Ar$^{+2}$(7751) & [O III] & $^{*}$9.71$\pm$2.46(-7) &  & [O III] & $^{*}$5.33$\pm$1.39(-7)  \\
Ar$^{+2}$ & wm & 1.15$\pm$0.26(-6) &  & wm & \nodata  \\
Ar$^{+3}$(4740) & [O III] & $^{*}$1.26$\pm$0.24(-7) &  & [O III] & $^{*}$1.50$\pm$0.28(-7) \\
icf(Ar) &  & 1.12$\pm$0.19 &  &  & 1.08$\pm$0.02  \\
\\
C$^{+2}$(4267) & [O III] & $^{*}$1.75$\pm$0.32(-3) &  & [O III] & \nodata  \\
icf(C) &  & 1.12$\pm$0.19 &  &  & \nodata \\
\\
Cl$^{+2}$ & [O III] & 8.20$\pm$3.52(-8) &  & [O III] & \nodata  \\
Cl$^{+2}$(5517) & [O III] & $^{*}$7.95$\pm$6.74(-8) &  & [O III] & \nodata  \\
Cl$^{+2}$(5537) & [O III] & $^{*}$8.31$\pm$2.26(-8) &  & [O III] & \nodata  \\
Cl$^{+2}$ & wm & 8.20$\pm$3.43(-8) &  & wm & \nodata  \\
icf(Cl) &  & 1.00 &  &  &  \nodata \\
\\
N$^{+}$(6584) & [N II] & $^{*}$7.30$\pm$5.25(-6) &  & [N II] & $^{*}$2.58$\pm$0.50(-6)  \\
N$^{+}$(6548) & [N II] & $^{*}$7.02$\pm$4.97(-6) &  & [N II] & $^{*}$2.40$\pm$0.47(-6)  \\
N$^{+}$(5755) & [N II] & $^{*}$7.30$\pm$5.25(-6) &  & [N II] & \nodata  \\
N$^{+}$ & wm & 7.23$\pm$5.17(-6) &  & wm & 2.54$\pm$0.47(-6)  \\
icf(N) &  & 9.40$\pm$12.84 &  &  & 13.4$\pm$2.53 \\
\\
Ne$^{+2}$(3869) & [O III] & $^{*}$7.94$\pm$2.00(-5) &  & [O III] & $^{*}$5.49$\pm$1.33(-5)  \\
Ne$^{+2}$(3967) & [O III] & 7.50$\pm$2.54(-5) &  & [O III] & 6.26$\pm$1.83(-5)  \\
icf(Ne) &  & 1.12 $\pm$0.20&  &  & 1.02$\pm$0.00  \\
\\
S$^{+}$ & [N II] & $^{*}$3.01$\pm$4.45(-7) &  & [N II] & $^{*}$1.77$\pm$0.33(-7)  \\
S$^{+}$(6716) & [N II] & 3.01$\pm$4.46(-7) &  & [N II] & 1.75$\pm$0.34(-7)  \\
S$^{+}$(6731) & [N II] & 3.01$\pm$4.44(-7) &  & [N II] & 1.79$\pm$0.35(-7)  \\
S$^{+}$ & [S II] & 3.25$\pm$10.8(-7) &  & [S II] & \nodata  \\
S$^{+2}$(6312) & [O III] & $^{*}$4.33$\pm$1.15(-6) &  & [O III] & $^{*}$2.93$\pm$0.81(-6)  \\
icf(S) &  & 1.31$\pm$0.34 &  &  & 2.03$\pm$0.18  \\
\\
\enddata

\end{deluxetable}

\clearpage

\begin{deluxetable}{lccc}
\tabletypesize{\small}
\setlength{\tabcolsep}{0.07in}
\tablecolumns{4}
\tablewidth{0.0in}
\tablenum{A3}
\tablecaption{Temperatures, Densities \& Total Abundances}
\tablehead{
\colhead{Parameter}&
\colhead{PN174}&
\colhead{PN2496}&
\colhead{Solar Reference\tablenotemark{a}}
}
\startdata 
T$_{[O III]}$ & 10280$\pm$573 &   12180$\pm$709&  \\
T$_{[N II]}$ & 11140$\pm$3197 &  12340$\pm$364\tablenotemark{b}&  \\
T$_{[O II]}$ & 11700$\pm$12430 & \nodata &\\
T$_{[S II]}$ & 10570$\pm$15560 &  \nodata & \\
Ne$_{[S II]}$ & 14770$\pm$21540&   15000\tablenotemark{c} &\\
\medskip
Ne$_{[Cl III]}$ & 16410$\pm$5557 &  \nodata &\\
He/H & 0.116$\pm$0.015 & 0.104$\pm$0.014 & 8.50(-2) \\
C/H & 1.96$\pm$0.47(-3) & \nodata & 2.69(-4)\\
C/O & 5.00$\pm$1.55 & \nodata & 0.550\\
N/H & 6.80$\pm$4.88(-5) & 3.40$\pm$0.75(-5) & 6.76(-5) \\
N/O & 0.173$\pm$0.166 & 0.128$\pm$0.021 & 0.138\\
O/H & 3.92$\pm$1.35(-4) & 2.66$\pm$0.63(-4) & 4.89(-4)\\
Ne/H & 8.86$\pm$3.17(-5) & 5.61$\pm$1.35(-5) & 8.51(-5)\\
Ne/O & 0.226$\pm$0.037 & 0.211$\pm$0.035 & 0.174\\
S/H & 6.05$\pm$1.80(-6) & 6.30$\pm$2.11(-6) & 1.32(-5)\\
S/O & 1.54$\pm$0.66(-2) & 2.37$\pm$0.49(-2) & 2.70(-2)\\
Cl/H & 8.20$\pm$3.43(-8) & \nodata & 3.16(-7)\\
Cl/O & 2.09$\pm$0.33(-4) & \nodata & 6.46(04)\\
Ar/H & 1.42$\pm$0.41(-6) & 7.39$\pm$1.68(-7) & 2.51(-6)\\
Ar/O & 3.63$\pm$0.74(-3) & 2.78$\pm$0.53(-3) & 5.13(-3)\\
\\
\enddata

\tablenotetext{a}{\citet{A09}}
\tablenotetext{b}{estimated using T$_{[O III]}$}
\tablenotetext{c}{high-density limit}

\end{deluxetable}
\clearpage

\begin{deluxetable}{lcccccc}
\tabletypesize{\small}
%\rotate
\tablecolumns{7}
\tablewidth{0pc}
\tablenum{A4}
\tablecaption{Derived Stellar Parameters}
\tablehead{
\colhead{Merrett \#} &
\colhead{log T$_{eff}$} &
\colhead{log L/L$_{\odot}$}&
\colhead{M$_f$/M$_{\odot}$\tablenotemark{a}} &
\colhead{L$_{5007}$/L$_*$} &
\colhead{M$_i$/M$_{\odot}$\tablenotemark{b}} &
\colhead{t$_{ms}$ (GY)\tablenotemark{c}} 
}
\startdata
M174	&	4.85	&	 3.70 &	0.597 &	0.05	&	1.82 &	1.53 \\
M2496	&	 4.86	& 	4.00 &	0.677 &	0.03	&	2.51 &	0.58 \\
\enddata
\tablenotetext{a}{ final stellar mass, M$_f$, from model tracks of \citet{VW94}}
\tablenotetext{b}{ initial stellar mass, M$_i$, determined from \citet{Ca08}}
\tablenotetext{c}{ main-sequence lifetime, t$_{ms}$, determined from results of \citet{Sc92}}

\end{deluxetable}


\begin{thebibliography}{}

\bibitem[Acker \& Neiner(2003)]{A03} Acker, A., \& Neiner, C.\ 2003, \aap, 403, 659
\bibitem[Asplund et al.(2009)]{A09} Asplund, M., Grevesse, N., Sauval, A.~J., \& Scott, P.\ 2009, \araa, 47, 481 
\bibitem[Azimlu et al.(2011)]{Az11} Azimlu, M., Marciniak, R., \& Barmby, P.\ 2011, \aj, 142, 139 
\bibitem[Barker et al.(2011)]{Bar11} Barker, M.~K., Ferguson, A.~M.~N., Cole, A.~A., et al.\ 2011, \mnras, 410, 504 
%\bibitem[Bastian et al.(2009)]{Bas09} Bastian, N., Trancho, G., Konstantopoulos, I.~S., \& Miller, B.~W.\ 2009, \apj, 701, 607 
\bibitem[Bernard et al.(2012)]{Ber12} Bernard, E.~J., Ferguson, A.~M.~N., Barker, M.~K., et al.\ 2012, \mnras, 420, 2625 (BFB12)
\bibitem[Bird et al.(2012)]{Bird12} Bird, J.~C., Kazantzidis, S., \& Weinberg, D.~H.\ 2012, \mnras, 420, 913 
\bibitem[Blair et al.(1982)]{B82} Blair, W.~P., Kirshner, R.~P., \& Chevalier, R.~A.\ 1982, \apj, 254, 50 
\bibitem[Braun \& Thilker(2004)]{BT04} Braun, R., \& Thilker, D.~A.\ 2004, \aap, 417, 421 
\bibitem[Braun et al.(2009)]{B09} Braun, R., Thilker, D.~A., Walterbos, R.~A.~M., \& Corbelli, E.\ 2009, \apj, 695, 937
\bibitem[Bresolin(2011)]{B11} Bresolin, F.\ 2011, \apj, 730, 129
\bibitem[Brown et al.(2008)]{BB08} Brown, T.~M., Beaton, R., Chiba, M., et al.\ 2008, \apjl, 685, L121 
\bibitem[Brown et al.(2006)]{BS06} Brown, T.~M., Smith, E., Ferguson, H.~C., et al.\ 2006, \apj, 652, 323
\bibitem[Catal{\'a}n et al.(2008)]{Ca08} Catal{\'a}n, S., Isern, J., Garc{\'{\i}}a-Berro, E., \& Ribas, I.\ 2008, \mnras, 387, 1693 
\bibitem[Chapman et al.(2006)]{Ch06} Chapman, S.~C., Ibata, R., Lewis, G.~F., et al.\ 2006, \apj, 653, 255
\bibitem[Chemin et al.(2009)]{Chem09} Chemin, L., Carignan, C., \& Foster, T.\ 2009, \apj, 705, 1395 
\bibitem[Choi et al.(2002)]{C02} Choi, P.~I., Guhathakurta, P., \& Johnston, K.~V.\ 2002, \aj, 124, 310
\bibitem[Ciardullo(2010)]{C10} Ciardullo, R.\ 2010, \pasa, 27, 149 
\bibitem[Ciardullo et al.(2005)]{C05} Ciardullo, R., Sigurdsson, S., Feldmeier, J.~J., \& Jacoby, G.~H.\ 2005, \apj, 629, 499 
\bibitem[Cioni(2009)]{Ci09} Cioni, M.-R.~L.\ 2009, \aap, 506, 1137 
\bibitem[Collins et al.(2011)]{Col11} Collins, M.~L.~M., Chapman, S.~C., Ibata, R.~A., et al.\ 2011, \mnras, 413, 1548
\bibitem[Corbelli et al.(2010)]{Corb10} Corbelli, E., Lorenzoni, S., Walterbos, R., Braun, R., \& Thilker, D.\ 2010, \aap, 511, A89
\bibitem[Courteau et al.(2011)]{Cour11}Courteau, S., Widrow,  L.~M., McDonald, M., et al.\ 2011, \apj, 739, 201
\bibitem[Cox \& Loeb(2008)]{2008MNRAS.386..461C} Cox, T.~J., \& Loeb, A.\ 2008, \mnras, 386, 461 
\bibitem[Dennefeld \& Kunth(1981)]{dk81} Dennefeld, M., \& Kunth, D. 1981, \aj, 86, 989
\bibitem[Esteban et al.(2004)]{E04} Esteban, C., Peimbert, M., Garc{\'i}a-Rojas, J., et al. 2004, MNRAS, 355, 229
\bibitem[Ferland et al.(1998)]{F98} Ferland, G.~J., Korista, K.~T., Verner, D.~A., et al.\ 1998, \pasp, 110, 761 
\bibitem[Freedman \& Madore(1990)]{FM90} Freedman, W.~L., \& Madore, B.~F.\ 1990, \apj, 365, 186
\bibitem[Gogarten et al.(2010)]{2010ApJ...712..858G} Gogarten, S.~M., Dalcanton, J.~J., Williams, B.~F., et al.\ 2010, \apj, 712, 858 
%\bibitem[Gon{\c c}alves et al.(2012)]{G12} Gon{\c c}alves, D.~R., Teodorescu, A.~M., Alves-Brito, A., M{\'e}ndez, R.~H., \& Magrini, L.\ 2012, \mnras, 425, 2557 
\bibitem[G{\'o}rny et al.(2009)]{G09} G{\'o}rny, S.~K., Chiappini, C., Stasi{\'n}ska, G., \& Cuisinier, F.\ 2009, \aap, 500, 1089 (Stas13) 
\bibitem[Hammer et al.(2010)]{Ham10} Hammer, F., Yang, Y.~B., Wang, J.~L., et al.\ 2010, \apj, 725, 542
%\bibitem[Haywood et al.(2013)]{2013arXiv1305.4663H} Haywood, M., Di Matteo, P., Lehnert, M., Katz, D., \& Gomez, A.\ 2013, arXiv:1305.4663 
\bibitem[Henry et al.(2010)]{H10} Henry, R.~B.~C., Kwitter, K.~B., Jaskot, A.~E., et al.\ 2010, \apj, 724, 748
\bibitem[Hibbard et al.(2001)]{Hib01} Hibbard, J.~E., van Gorkom, J.~H., Rupen, M.~P., \& Schiminovich, D.\ 2001, Gas and Galaxy Evolution, 240, 657
\bibitem[Ibata et al.(2005)]{Ib05} Ibata, R., Chapman, S., Ferguson, A.~M.~N., et al.\ 2005, \apj, 634, 287
\bibitem[Irwin et al.(2005)]{I05} Irwin, M.~J., Ferguson, A.~M.~N., Ibata, R.~A., Lewis, G.~F., \& Tanvir, N.~R.\ 2005, \apjl, 628, L105
\bibitem[Jacoby \& Ciardullo(1999)]{JC99} Jacoby, G.H., \& Ciardullo, R.B. 1999, \apj, 515, 169
\bibitem[Kalirai et al.(2006)]{K06} Kalirai, J.~S.,  Gilbert, K.~M., Guhathakurta, P., et al.\ 2006, \apj, 648, 389 
\bibitem[Karakas(2010)]{K10} Karakas, A.~I.\ 2010, \mnras, 403, 1413 
\bibitem[Kennicutt(1988)]{K98} Kennicutt, R.~C., Jr.\ 1998, \apj, 498, 541 
\bibitem[Kwitter \& Henry(2001)]{KH01} Kwitter, K.~B., \& Henry, R.~B.~C.\ 2001, \apj, 562, 804 
\bibitem[Kwitter \& Henry(2012)]{KH12} Kwitter, K.~B., \& Henry, R.~B.~C.\ 2012, IAU Symposium, 283, 119 
\bibitem[Kwitter et al.(2012)]{KLBH12} Kwitter, K.~B., Lehman, E.~M.~M., Balick, B., \& Henry, R.~B.~C.\ 2012, \apj, 753, 12 (KL12; Paper~I)
\bibitem[Kwitter et al.(2013)]{KLBH13} Kwitter, K.~B., Lehman, E.~M.~M., Balick, B., \& Henry, R.~B.~C.\ 2013, \apj, 768, 97 (KL13)
\bibitem[Lewis et al.(2013)]{LB13} Lewis, G.~F., Braun, R., McConnachie, A.~W., et al.\ 2013, \apj, 763, 4 
\bibitem[Lockman et al.(2012)]{LF12} Lockman, F.~J., Free, N.~L., \& Shields, J.~C.\ 2012, \aj, 144, 52 
\bibitem[Loebman et al.(2011)]{L11} Loebman, S.~R., Ro{\v s}kar, R., Debattista, V.~P., et al.\ 2011, \apj, 737, 8 
\bibitem[L{\'o}pez-S{\'a}nchez \& Esteban(2010)]{LS10} L{\'o}pez-S{\'a}nchez, {\'A}.~R., \& Esteban, C.\ 2010, \aap, 517, A85 
\bibitem[Magrini et al.(2009)]{Mag09} Magrini, L., Stanghellini, L., \& Villaver, E.\ 2009, \apj, 696, 729
\bibitem[Marigo(2001)]{Mar01} Marigo, P.\ 2001, \aap, 370, 194
\bibitem[Marigo et al.(2004)]{Mar04} Marigo, P., Girardi, L., Weiss, A., Groenewegen, M.~A.~T., \& Chiosi, C.\ 2004, \aap, 423, 995 
\bibitem[McConnachie et al.(2009)]{Mc09} McConnachie, A.~W., Irwin, M.~J., Ibata, R.~A., et al.\ 2009, \nat, 461, 66 
\bibitem[M{\'e}ndez et al.(2008)]{Men08} M{\'e}ndez, R.~H., Teodorescu, A.~M., Sch{\"o}nberner, D., Jacob, R., \& Steffen, M.\ 2008, \apj, 681, 325 
\bibitem[Merrett et al.(2006)]{m06}Merrett, H.R., Merrifield, M.R.,  Douglas, N.G., Kuijken, K., Romanowsky, A.J., Napolitano, N.R., Arnaboldi, M., Capaccioli, M., Freeman, K.C., Gerhard, O., Coccato, L., Carter, D., Evans, N.W., Wilkinson, M.I., Halliday, C., \& Bridges, T.J. 2006, \mnras, 369, 120 (M06)
\bibitem[Monk et al.(1988)]{M98} Monk, D.~J., Barlow, M.~J., \& Clegg, R.~E.~S.\ 1988, \mnras, 234, 583 
\bibitem[Montalto et al.(2009)]{Mon09} Montalto, M., Seitz, S., Riffeser, A., et al.\ 2009, \aap, 507, 283 
\bibitem[Pilyugin et al.(2006)]{P06} Pilyugin, L.~S., V{\'{\i}}lchez, J.~M., \& Thuan, T.~X.\ 2006, \mnras, 370, 1928 
\bibitem[Purcell et al.(2010)]{P10} Purcell, C.~W., Bullock, J.~S., \& Kazantzidis, S.\ 2010, \mnras, 404, 1711 
\bibitem[Richardson et al.(2008)]{Ri08} Richardson, J.~C., Ferguson, A.~M.~N., Johnson, R.~A., et al.\ 2008, \aj, 135, 1998 
\bibitem[Richardson et al.(2009)]{Ri09} Richardson, J.~C., Ferguson, A.~M.~N., Mackey, A.~D., et al.\ 2009, \mnras, 396, 1842 
\bibitem[Richer \& McCall(2008)]{RM08} Richer, M.~G., \& McCall, M.~L.\ 2008, \apj, 684, 1190
\bibitem[Richer et al.(1997)]{RM97} Richer, M.~G., McCall, M.~L., \& Arimoto, N.\ 1997, \aaps, 122, 215
%\bibitem[Sabbi et al.(2009)]{2009ApJ...703..721S} Sabbi, E., Gallagher, J.~S., Tosi, M., et al.\ 2009, \apj, 703, 721 
\bibitem[Sales et al.(2009)]{Sal09} Sales, L.~V., Helmi, A., Abadi, M.~G., et al.\ 2009, \mnras, 400, L61 
\bibitem[Sanders et al.(2012)]{SC12} Sanders, N.~E., Caldwell, N., McDowell, J., \& Harding, P.\ 2012, \apj, 758, 133 (SC12)
\bibitem[Savage \& Mathis(1979)]{SM79} Savage, B.~D., \& Mathis, J.~S.\ 1979, \araa, 17, 73 
\bibitem[Schaller et al.(1992)]{Sc92} Schaller, G., Schaerer, D., Meynet, G., \& Maeder, A.\ 1992, \aaps, 96, 269 
\bibitem[Sch{\"o}nberner et al.(2005)]{S05} Sch{\"o}nberner, D., Jacob, R., Steffen, M., et al.\ 2005, \aap, 431, 963 
\bibitem[Sch{\"o}nberner et al.(2007)]{S07} Sch{\"o}nberner, D., Jacob, R., Steffen, M., \& Sandin, C.\ 2007, \aap, 473, 467
\bibitem[Sch{\"o}nrich \& Binney(2009a)]{ScB09} Sch{\"o}nrich, R., \& Binney, J.\ 2009, \mnras, 396, 203
\bibitem[Sch{\"o}nrich \& Binney(2009b)]{ScB092} Sch{\"o}nrich, R., \& Binney, J.\ 2009, \mnras, 399, 1145 
\bibitem[Sellwood \& Binney(2002)]{SB02} Sellwood, J.~A., \& Binney, J.~J.\ 2002, \mnras, 336, 785 
%\bibitem[Shaw et al.(2010)]{SLS10} Shaw, R.~A., Lee, T.-H., Stanghellini, L., et al.\ 2010, \apj, 717, 562 
%\bibitem[Stanghellini et al.(2009)]{2009ApJ...702..733S} Stanghellini, L., Lee, T.-H., Shaw, R.~A., Balick, B., \& Villaver, E.\ 2009, \apj, 702, 733
\bibitem[Spitoni et al.(2013)]{SM13} Spitoni, E., Matteucci, F., \& Marcon-Uchida, M.~M.\ 2013, \aap, 551, A123 
\bibitem[Stanghellini et al.(2010)]{Stang10} Stanghellini, L., Magrini, L., Villaver, E., \& Galli, D.\ 2010, \aap, 521, A3
\bibitem[Stasi{\'n}ska et al.(2013)]{Stas13} Stasi{\'n}ska, G., Pe{\~n}a, M., Bresolin, F., \& Tsamis, Y.~G.\ 2013, \aap, 552, A12 (Stas13)
\bibitem[Thilker et al.(2004)]{T04} Thilker, D.~A., Braun, R., Walterbos, R.~A.~M., et al.\ 2004, \apjl, 601, L39 
\bibitem[Trundle et al.(2002)] {T02} Trundle, C., Dufton, P.L., Lennon, D.J., Smartt, S.J., \& Urbaneja, M.A. 2002, \aap, 395, 519
\bibitem[Vassiliadis \& Wood(1994)]{VW94} Vassiliadis, E., \& Wood, P.~R.\ 1994, \apjs, 92, 125
%\bibitem[de Vaucouleurs et al.(1991)]{deV91} de Vaucouleurs, G., de Vaucouleurs, A., Corwin, H.~G., Jr., et al.\ 1991, Third Reference Catalogue of Bright Galaxies. Volume II: Data for galaxies between 0$^{h}$ and 12$^{h}$.~Springer, New York, NY (USA), 1991
\bibitem[Werk et al.(2011)]{W11} Werk, J.~K., Putman, M.~E., Meurer, G.~R., \& Santiago-Figueroa, N.\ 2011, \apj, 735, 71 
\bibitem[Zurita \& Bresolin(2012)]{ZB12} Zurita, A., \& Bresolin, F.\ 2012, \mnras, 427, 1463  Bresolin, F.\ 2012, \mnras, 427, 1463 


\end{thebibliography}
\end{document}